\pdfoutput=1
\documentclass[11pt]{article}

\usepackage[]{acl}

\usepackage{times}
\usepackage{latexsym}
\usepackage{amsfonts,amsmath,amssymb}
\usepackage{multirow}

\usepackage{subcaption} % For subfigures

\usepackage[T1]{fontenc}
\usepackage[utf8]{inputenc}

\usepackage{microtype}

\usepackage{inconsolata}

\usepackage{graphicx}
\newcommand\scalemath[2]{\scalebox{#1}{\mbox{\ensuremath{\displaystyle #2}}}}

\title{QABISAR: Query-Article Bipartite Interactions for Statutory Article Retrieval}

\author{Santosh T.Y.S.S, Hassan Sarwat, Matthias Grabmair \\
School of Computation, Information, and Technology; \\
Technical University of Munich, Germany \\
\texttt{\{santosh.tokala, hassan.sarwat, matthias.grabmair\}@tum.de}
}

\begin{document}
\maketitle
\begin{abstract}
In this paper, we introduce QABISAR, a novel framework for statutory article retrieval, to overcome the semantic mismatch problem when modeling each query-article pair in isolation, making it hard to learn representation that can effectively capture multi-faceted information. QABISAR leverages bipartite interactions between queries and articles to capture diverse aspects inherent in them. Further, we employ knowledge distillation to transfer enriched query representations from the graph network into the query bi-encoder, to capture the rich semantics present in the graph representations, despite absence of graph-based supervision for unseen queries during inference. Our experiments on a real-world expert-annotated dataset demonstrate its effectiveness.
\end{abstract}

\section{Introduction}
In an age where legal complexities are challenging for many individuals, there is a pressing need to bridge the gap between legal expertise and public understanding \cite{ponce2019global}. One critical step in this process is Statutory Article Retrieval (SAR), which involves identifying relevant statutes for a legal question. SAR plays a vital role in providing initial legal assistance by offering foundational insights into the law. Beyond statutes, previous works have explored retrieval of similar prior cases \cite{goebel2023summary, santosh2024ecthr} or pertinent information from case documents in response to legal queries \cite{santosh2024query}. Leveraging advanced technologies for SAR can enhance the accuracy of legal insights, ultimately making legal knowledge more accessible and understandable to a broader audience.

Traditionally, SAR methods have been explored using the COLIEE Statute Law Corpus \cite{rabelo2021coliee}, containing questions linked to relevant articles from the Japanese Civil Code. However, these questions which are obtained from legal bar exam yes or no questions, are quite different from those posed by ordinary citizens, often being vague and underspecified. To address this, \citet{louis2022statutory} developed the Belgian Statutory Article Retrieval Dataset (BSARD), featuring french legal questions from Belgian citizens labeled by legal experts with references to relevant articles from Belgian legislation, which we use in our study.

Traditional SAR techniques included BM25, TF-IDF \cite{yoshioka2018overview}, Indri \cite{strohman2005indri} and Word Movers’ Distance \cite{kusner2015word}. With the rise of pre-trained models, BERT and their ensembles have become popular \cite{kim2019statute, rabelo2021coliee, rabelo2022overview}. Recently, dense retrieval methods have gained attention \cite{louis2022statutory} and were enhanced further through synthetic query generation
and legal domain-oriented pre-training \cite{louis2023finding}. Additionally, \citet{louis2023finding} has demonstrated that articles are not completely independent of each other but an ensemble of interdependent rules organized into different codes, books, titles, chapters, and sections. They leveraged the hierarchical organization of statute law and utilized graph neural networks to enrich article representations by exploiting the interdependencies among articles within the topological structure. Orthogonal to these improvements, \citet{santosh2024cusines} introduced a curriculum-based negative sampling strategy to make the model distinguish easier negatives in the initial stages of learning and progressively tackle more difficult ones. 

Existing SAR works primarily focus on capturing the semantic relevance between individual query and article pairs in isolation. However, we argue that this approach may lead to sub-optimal representations, particularly in scenarios where both queries and articles contain multifaceted information. Articles, for instance, can cover a variety of semantics relevant to different queries, while each query may require multiple relevant articles to comprehensively address its various aspects. Recognizing this inherent many-to-many relationship, our work takes a different approach. We leverage a query-article bipartite graph where nodes represent either a query or an article, and edges between them signify their relevance. We further augment this bipartite graph with hierarchical organization of statutes, with additional structural links such as sections, chapters, titles etc, to facilitate cross-article dependencies through their neighbourhood hops \cite{louis2023finding}. We employ a Graph Attention Network on this augmented graph to aggregate information across the graph, allowing to capture the multiple interactions between queries and articles simultaneously, leading to effective capture of the diverse aspects inherent in each of them.

During inference, we face the challenge of utilizing only article representations from the graph network and fall back on query encoder to obtain query embeddings, as unseen queries at test time are absent in the constructed graph. To ensure that the query encoder representations are as expressive as the graph representations, we adopt knowledge distillation (KD) \cite{hinton2015distilling} which aims to improve student models with the aid of teacher models which usually have same architecture with greater number of layers and dimensions \cite{wang2020minilm}. KD has been explored in IR tasks earlier, different from conventional setting, where they employ more expressive cross-encoder models as teachers and bi-encoders as students \cite{qu2021rocketqa,lu2022ernie,hofstatter2020improving,choi2021improving}. In this work, we use KD to facilitate representation transfer of queries from the graph network to the query encoder. By doing so, we aim to equip the query encoder with the ability to capture the rich semantics present in the graph representations, despite the absence of explicit graph-based supervision during inference.

We apply our approach, QABISAR, on the publicly available BSARD dataset \cite{louis2022statutory}, demonstrating the effectiveness of leveraging bipartite interactions between queries and articles, as well as knowledge distillation for representational transfer.

\section{Our Method: QABISAR}
\noindent \textbf{Statutory Article Retrieval:} Given a question $q$ and corpus of statues $P =\{p_1, p_2, \ldots, p_m\}$, the task of SAR is to retrieve a smaller set of statutes $P_q$ ($|P_q|<< |P|)$ ranked in terms of their relevancy to answer the query. We mainly deal with optimizing the recall of the SAR system acting as pre-fetcher component, leaving the re-ranker component optimized for precision, for future. 

QABISAR involves two stages of training. The first stage employs a dense bi-encoder which maps query and article into representations independently, while the second stage utilizes a graph encoder, designed to capture the bipartite interactions between queries and articles, enhancing retrieval by learning multi-faceted representations.

\subsection{Dense bi-encoder} 
We use a dual-encoder architecture \cite{karpukhin2020dense} with query and statute encoder mapping each of them into a k-dimensional vector and the relevance score is computed using dot product between the encodings of query $q$ and statute $p_i$ as 
%\begin{equation}
    $f(q,p_i)= E_q(q) \cdot E_p({p_i})$
%\end{equation}
where $E_q$, $E_p$ denote query and statute encoder. We use BERT-based model \cite{devlin2018bert} as query encoder to obtain query embedding from [CLS] representation. To account for longer length of articles, we use hierarchical article encoder \cite{pappagari2019hierarchical} where the article is split into different chunks and each of them is independently encoded using a BERT-based model and then [CLS] representations from each chunk along with learnable position encodings are passed into a transformer which are then max-pooled to obtain the article embedding.

Dense bi-encoder module is trained with contrastive loss whose objective is to pull the representations of the query $q$ and relevant articles $S_q$ together (as positives), while pushing apart irrelevant ones $P_q' = P - P_q$ (as negatives). However, %the irrelevant statutes for a query form a large pool of candidates, which would make 
training with all the negatives is inscalable given larger corpus. To alleviate this issue, negative sampling has been employed where some irrelevant documents are sampled for each query during training making the final objective function as follows:
\begin{equation*}
\scalemath{0.8}{
    L(q, P_q, P_q') = \sum_{p \in P_q}  -log  \frac{\exp(f(q,p)/\tau)}{ \sum_{c \in \{p\} \cup P_q'} \exp(f(q,c)/\tau)}}
\end{equation*}
where hyperparameter $\tau$ is a scalar temperature. Following \citet{karpukhin2020dense}, we consider two types of negatives: (i) in-batch -articles paired with the other queries in the the same batch, and (ii) BM25- top articles returned by BM25 that are not relevant to the query.

\subsection{Graph Encoder}
%While the bi-encoder effectively captures the semantic relevance between individual query and article pair independently, we posit that the learned representations may be sub-optimal for capturing the multiple aspects inherent in both queries and articles. Each article can encompass multiple semantics relevant to different queries, and conversely, each query may require multiple relevant articles to fully address its various aspects. 
To capture many-to-many interactions between query and articles effectively, we construct a query-article bipartite graph from training data and utilize graph attention network, to enrich the representations of both queries and articles through their multiple interactions simultaneously. %, thereby effectively capturing the diverse aspects in each of them.

\noindent \emph{Graph Construction:} 
We construct a bipartite graph utilizing all queries from the training set and all the articles from the corpus as nodes, establishing edges between queries and their corresponding relevant articles. Additionally, we augment our bipartite graph with the hierarchical organizational topology of statute structure by introducing additional nodes to represent sections, chapters, titles, and books and creating edges to denote the hierarchical connections between these structural units and the article nodes. This enables the article nodes to learn the complementary information from neighbouring elements as articles are not completely independent of each other but an ensemble of interdependent rules organized into different codes, books, titles, chapters, and sections \cite{louis2023finding}. We also label edges based on the type of nodes they connect (i.e., Query-Article, Section-Article etc). 

\noindent \emph{Node Initialization:} We use the article encoder to obtain embeddings for structural units in the legislative topology graph such as section, article nodes and query encoder to obtain embeddings for query nodes based on their textual content.

\noindent \emph{Graph Attention Network (GAT):} A graph neural network layer updates every node representation by aggregating its neighbors representations. GAT \cite{velivckovic2018graph,brody2021attentive} inject the graph structure into the attention mechanism by performing masked attention using neighborhood of a node. It employs a  K multi-headed attention mechanism with residual connections and then their features are
concatenated, resulting in the updated node feature representation as follows:
\begin{equation*}
\scalemath{0.8}{
    \mathbf{x_i'} =  \big\Vert_{k=1}^K \sigma \left(\alpha_{i,i}^k \mathbf{W^k_s} \mathbf{x_i} + \sum_{j \in N(i)} \left( \alpha_{i,j}^k \mathbf{W^k_t} \mathbf{x_j} \right)\right)}
\end{equation*}
where $\big\Vert$ indicates concatenation, $\sigma$ indicates non-linearity, $W_s, W_t$ indicates learnable weight matrices.  $\alpha_{i,j}^k$ denote attention weight computed using the node features $x_i, x_j$ and edge connecting them $e_{i,j}$, which in our case indicate the node types.
\begin{equation*}
\scalemath{0.8}{
    \mathbf{\alpha_{i,j}^k} = \text{Softmax}\left((\mathbf{a}^k)^T \text{LeakyReLU}\left(\left[W^k_s \mathbf{x_i} \, || \, W^k_t \mathbf{x_j} \, || \, W^k_e \mathbf{e_{i,j}}\right]\right)\right)}
\end{equation*}

To learn GAT parameters, we adopt the same contrastive learning used to train bi-encoder by obtaining the query and article representations from the graph nodes. We only sample sub-graph with L-hop neighbours (L denote number of GAT layers) based on the current batch of query, articles to save computational cost and pass them into GAT to extract node features for loss computation. These bi-partite interactions, lead to enriched representations for both the articles and queries representations capturing multiple aspects covered in them, However, during inference, we can not use query representations from graph as we encounter   unseen queries that are not present in the constructed graph, resulting in an inductive learning setting on graphs. 

We employ knowledge distillation \cite{hinton2015distilling} to facilitate representation transfer for queries from the graph network to the query encoder. This involves distilling the relevance scores assessed with the query representations from the graph (acting as the teacher model) into the query bi-encoder (student model). Formally, given a query q and a list of candidate articles P =$\{p_1, p_2, \ldots, p_m\}$, we obtain article representations from graph network $p^g$ and query representations from query bi-encoder $q^b$ and graph network $q^g$. We compute two relevance scores between query and article representations and convert them into probability distributions of the scores over candidate articles. We apply Knowledge distillation to mimic the scoring distribution of a more expressive graph network based relevance scores $\text{s}(q^g, p^g)$ with the distribution of a bi-encoder $\text{s}(q^b, p^g)$ measured by KL divergence as follows
\begin{align*}
    L_{KD} &= \sum_{q \in Q, p \in P} \text{s}(q^g, p^g) \cdot \log \frac{\text{s}(q^g, p^g)}{\text{s}(q^b, p^g)} \\
    & \text{s}(q^g, p^g) = \frac{e^{f(q^g,p^g)}}{\sum_{p' \in P} e^{f(q^g,p'^g)}}
\end{align*}
Our final loss for the second-stage training is the sum of contrastive loss using graph representations and the knowledge distillation loss. This joint training drives the bi-encoder to update in tandem with the graph representations, which are also initialized with bi-encoder representations during the start.

\section{Experiments}
\subsection{Dataset \& Baselines}
We use BSARD \cite{louis2022statutory} containing  1108 french legal questions, with references to relevant articles from a corpus of 22,600 Belgian legal articles. We derive following baselines from prior works of \citet{louis2022statutory,louis2023finding}: Sparse retriever such as (i) BM25 \cite{robertson1995okapi}, Dense retrievers such as (ii) Bi-encoder with LegalCamemBERT as query and article encoder (BE w/o Hier.) (iii) Bi-encoder with hierarchical variant for article encoding (BE) (iv) BE along with graph encoder applied on statute structure graph (BE+GE-Stat.) Implementation details can be found in App. \ref{impl}.

\begin{table}[]
\scalebox{0.9}{
\begin{tabular}{|l|ccc|c|c|}
\hline
\multicolumn{1}{|c|}{\textbf{}} & \multicolumn{3}{c|}{\textbf{R@}}                                                     & \multirow{2}{*}{\textbf{MAP}} & \multirow{2}{*}{\textbf{MRP}} \\ \cline{1-4}
\textbf{Method}                 & \multicolumn{1}{c|}{\textbf{100}} & \multicolumn{1}{c|}{\textbf{200}} & \textbf{500} &                               &                               \\ \hline
BM25                            & \multicolumn{1}{c|}{49.3}         & \multicolumn{1}{c|}{57.3}         & 63.0         & 16.8                          & 13.6                          \\ \hline
BE w/o Hier.                    & \multicolumn{1}{c|}{77.9}         & \multicolumn{1}{c|}{81.8}         & 88.1         & 36.4                          & 30.1                          \\ \hline
BE                              & \multicolumn{1}{c|}{79.9}         & \multicolumn{1}{c|}{83.3}         & 88.7         & 39.2                          & 31.6                          \\ \hline
BE+GE-Stat.                     & \multicolumn{1}{c|}{82.3}         & \multicolumn{1}{c|}{85.1}         & 89.9         & 42.6                          & 34.8                          \\ \hline
QABISAR                         & \multicolumn{1}{c|}{83.7}         & \multicolumn{1}{c|}{87.9}         & 91.3         & 43.1                          & 35.6                          \\ \hline
\end{tabular}}
\caption{Comparison of QABISAR with prior works.}
\label{tab:results}
\end{table}

\subsection{Performance comparison}
Following previous work, we evaluate the retriever's performance using Recall@k (R@K) (k=100,200,500), Mean Average Precision (MAP) and Mean R-Precision (MRP). R@K measures the proportion of relevant articles in the top k candidates, with results averaged across all instances. MAP and MRP provide the mean of average precision and R-Precision scores for each query where average precision is the average of Precision@k scores for every rank position of each relevant document and Precision@k represents the proportion of relevant documents in the top k candidates. R-Precision indicates proportion of the relevant articles in the top-k ranked ones where k is exact number of relevant articles for that query. Higher scores in these metrics indicate better performance. 

From Table \ref{tab:results}, QABISAR consistently outperforms prior works across all metrics. This validates the superiority of two key aspects: (a) enriched multi-faceted representations of articles  shaped by interactions with queries, going beyond BE+GE-Stat. which only considers cross-article dependencies from statute topology graph. (b) transfer of enriched query representations from the graph network to the query bi-encoder, enabling it to grasp the complex semantics embedded, even without direct graph-based guidance during inference.
\vspace{0.3em}

\begin{figure}[t]
\begin{minipage}[t]{0.22\textwidth}
\includegraphics[width=1.1\textwidth]{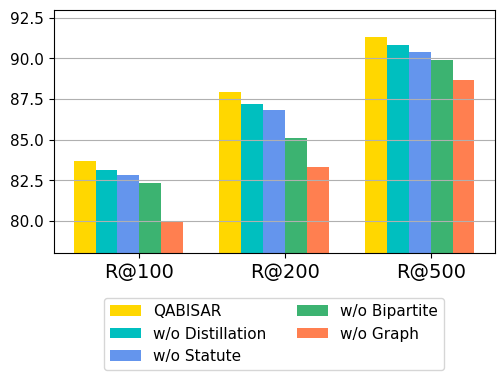}
\captionsetup{justification=centering,font=small}
\caption{Ablation Study on QABISAR}
\label{fig:ablation}
\end{minipage}\hfill
\begin{minipage}[t]{0.25\textwidth}
 \includegraphics[width=\textwidth]{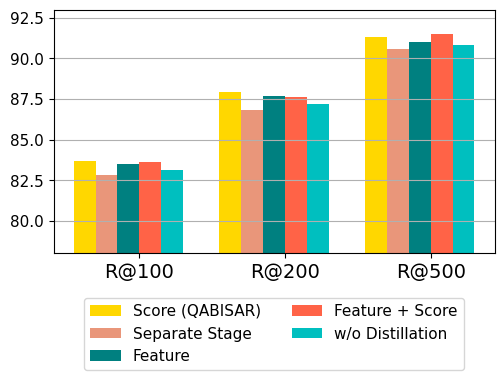}
\captionsetup{justification=centering,font=small}
\caption{Effect of various \\ distillation strategies}
\label{fig:dist}
\end{minipage}
\end{figure}

\noindent \textbf{Ablation Study on QABISAR:} We examine the effectiveness of each component in QABISAR through these ablations: (1) removing the distillation KD loss during training; (2) excluding bipartite graph interactions and using only the statute topology graph; (3) eliminating statute topology graph and relying solely on the bipartite graph; (4) removing the entire graph encoder. From Fig. \ref{fig:ablation}, we observe that distillation enables effective transfer of query representations into the bi-encoder, rendering them as expressive as graph representations. The inclusion of both the statute structure graph and the statute-query bipartite graph proves more effective, indicating the limitations of modeling relevance of each query-article pair in isolation. Between these two graph views, removing bipartite interactions has a more impact on performance, suggesting that they facilitate the effective capture of multi-faceted information present in articles by leveraging their simultaneous interactions with different queries, as well as other related articles connected by 1-hop neighbor bridges via queries.

\vspace{0.3em}

\noindent \textbf{Effect of Distillation:} We investigate various strategies for distilling query representations from the graph into the bi-encoder: (i) utilizing relevance score via KL divergence loss (ii) feature distillation through $L_2$ distance loss between both the query representations  \cite{heo2019comprehensive} (iii) combining both feature and score distillation. From Fig. \ref{fig:dist}, we observe that employing either or both distillation methods proves effective for transferring query representations compared to not using KD. Score distillation outperforms feature distillation, likely due to the limited number of queries in the training set, leading to overfitting losing generalizability with feature distillation alone. Moreover, score distillation enables the representations to become as expressive as graph representations for computing relevance scores, rather than aiming for exact replication as in feature distillation with $L_2$ loss. The combination of both methods does not yield as effective results as using score distillation alone, reinforcing overfitting effect. We explore the impact of joint training in QABISAR by applying KD after training the graph encoder separately. It underperforms compared to training without KD, as joint training allows for a more gradual steering of representations through the training, compared to the abrupt adjustment in separate stages.

\section{Conclusion}
We introduced QABISAR, a novel framework for SAR leveraging bipartite interactions between queries and articles to facilitate learning of multi-faceted representations and employ knowledge distillation to transfer the enriched query representations from graph into the bi-encoder. Through comprehensive ablation studies over publicly available BSARD dataset, we demonstrated its effectiveness. We hope this work to inspire more investigations on the different interaction schemes for capturing many-to-many relationship effectively.

\section*{Limitations}
Our experimental contributions are centered around the BSARD dataset, which is based on the French language and built on the Belgian legal system. It's important to note that the BSARD dataset introduces a linguistic bias as Belgium is a multilingual country with French, Dutch, and German speakers, yet the legal questions and provisions provided are only available in French \cite{louis2022statutory}. While our approach, QABISAR, demonstrates promising results within this specific context, its performance may vary when applied to legal jurisdictions with different legislative structures and languages. Expanding the application of QABISAR to other jurisdictions remains an avenue for future exploration, highlighting the need for concerted efforts to construct SAR datasets from diverse legal systems.

Our work primarily focuses on optimizing recall in the first stage of the retrieval system. For practical utility, a complementary re-ranking component is necessary to improve precision by identifying the most relevant statutes for each query, which are subsequently fed into a QA system to answer legal queries posed by individuals. Moreover, for this QA system to be truly accessible, it should not only retrieve relevant articles but also possess the capability to simplify legal texts, making them comprehensible to laypeople.

\section*{Ethics Statement}
We conduct experiments using the publicly available SAR dataset, BSARD \cite{louis2022statutory}. While leveraging pre-trained encoders enhances our model's performance, we acknowledge the inherent risk of inheriting biases embedded within these encoders. Consequently, it is imperative to subject our models to thorough scrutiny, particularly concerning equal treatment imperatives related to their performance, behavior, and intended use. Moreover, we recognize the potential for false information through such automated systems, which can have profound implications. Therefore, one needs to remain vigilant in consuming information from automated systems. Additionally, we are mindful of the broader societal impact of our technology, particularly its influence on marginalized communities. We advocate actively engaging with legal stakeholders to ensure ethical and responsible development and deployment of any SAR system. 

\bibliography{custom}

\appendix

\section{Implementation Details}
\label{impl}
Following \citet{louis2023finding}, we initialize the second-level encoder in the hierarchical article encoder with a two-layer transformer encoder featuring a hidden dimension of 768, an intermediate dimension of 3072, 12 heads, a dropout rate of 0.1, and the GeLU non-linearity function. Our training process for the dense encoder spans 15 epochs with a batch size of 24, employing the AdamW optimizer \cite{loshchilov2018decoupled} with hyperparameters $\beta_1$ = 0.9, $\beta_2$ = 0.999, $\epsilon$ = 1e-7, a weight decay of 0.01, and a learning rate warm-up for the first 5\% of training steps, reaching a maximum value of 2e-5, after which linear decay is applied. For the graph encoder, we conduct 20 epochs of training with a batch size of 512, using the AdamW optimizer with a learning rate of 2e-4. We use a learnable embedding layer for  edge types with dimension equal to node features. We assign weights of 0.7 and 0.3 to the contrastive loss and KD loss, respectively, in the second stage training.

\end{document}